\documentclass[10pt,amssymb,amsmath,aps,floatfix,prl]{revtex4}

\begin{document}

\title{Lorentz and CPT violations in Finsler spacetime}

\author{Zhe Chang$^{1,2}$\footnote{E-mail: changz@ihep.ac.cn}}
\author{Sai Wang$^{1}$\footnote{E-mail: wangsai@ihep.ac.cn}\footnote{E-mail: saiwangihep@gmail.com}
\footnote{Corresponding author at IHEP, CAS, 100049 Beijing, China}}
\affiliation{${}^1$\small{Institute of High Energy Physics\\Chinese Academy of Sciences, 100049 Beijing, China}\\
${}^2$\small{Theoretical Physics Center for Science Facilities\\ Chinese Academy of Sciences, 100049 Beijing, China}}

\begin{abstract}
Standard model with intrinsic Lorentz and CPT violations is proposed in a Finsler geometric framework. We present explicitly Lorentz and CPT--breaking Lagrangians of the matter fields and the gauge fields in locally Minkowski spacetime. The Lorentz invariance violation is found to be originated from the spacetime background deviating from the Minkowski one. Similarly, the CPT violation is determined by the behaviors of the locally Minkowski metric under the parity and time reversal operations. To help understanding phenomenologies, we compare the Finslerian model with the standard--model extension (SME) term by term at a first order approximation.
\end{abstract}

\maketitle

\section{1. Introduction}
\label{Introduction}
The standard model (SM) of particles has been a paradigm of the modern elementary particle physics.
Einstein's special relativity (SR) is one significant foundation of the SM.
Furthermore, the Lorentz--invariant SM preserves the CPT symmetry.
Therefore, it is significant to test the fate of the Lorentz and CPT violations via both theories and experiments.
In the past two decades, the Lorentz and CPT violations have acquired considerable amounts of investigations.
There have been several models that predict deformations or violations of the Lorentz and CPT symmetries.
The first one is the standard--model extension (SME) \cite{SME01,SME02}.
The SME originated from the concept of spontaneous breaking of the Lorentz and CPT symmetries in the string theory \cite{SBLV}.
It involves nonzero vacuum expectation values (vev) of tensor fields with spacetime indices.
The nonzero vev point out certain preferred directions of the spacetime.
The second model with Lorentz and CPT violations refers to
the spacetime foam model \cite{recoiling D-branes01,recoiling D-branes02,recoiling D-branes03}.
The foamy structure of the quantum--gravity (QG) gives rise to non--trivial optical properties for the vacuum.
This means a non--trivial vacuum refractive index for photons.
The third model is called deformed special relativity (DSR) \cite{DSR01,DSR02,DSR03,DSR04,DSR05}.
The DSR involves two invariant parameters, the speed of light and the Planck scale.
This character implies that the QG fluctuations induce modified dispersion relations (MDR) for particles in the vacuum \cite{MDR in DSR}.
The last model refers to very special relativity (VSR) \cite{VSR}.
The symmetry groups of the VSR include certain subgroups of the Poincare group,
which comprise the translation group and proper subgroups of the Lorentz group.

As the Lorentz symmetry resides in Minkowski spacetime,
the Lorentz violation may result from certain deviations of the spacetime from Minkowski one.
Actually, all the Lorentz--breaking models mentioned above have close connections with
Finsler geometry \cite{Book by Rund,Book by Bao,Book by Shen}.
In the SME, the nonzero vev denote the Lorentz and CPT--breaking coupling constants in the Lagrangians \cite{SME01,SME02}.
They characterize the anisotropy of the spacetime.
Recently, they were found to be related with certain fixed preferred directions in
Finsler spacetime \cite{Kostelecky_Finsler,SME Bogoslovsky01,SME Bogoslovsky02}.
The dispersion relations for quantum wave packets were related to a classical point--like Lagrangian in the form of Finsler geometry \cite{Kostelecky_Finsler}.
In the D-brane model of spacetime foam \cite{recoiling D-branes01,recoiling D-branes02,recoiling D-branes03},
D--particles are recoiled by photons.
Thus the spacetime metric, experienced by photons, is modified.
It was found that this metric belongs to Finsler geometry \cite{recoiling D-branes02}.
The DSR was also incorporated into the framework of Finsler geometry \cite{DSR in Finsler}.
The MDR of the DSR could be associated with certain Finsler line elements.
These Finsler structures refer to the so--called ``rainbow metric'' \cite{Rainbow metric}.
In addition, it has been shown that the Finsler structure \(d\tau=\left(\eta_{\mu\nu}dx^{\mu}dx^{\nu}\right)^{\frac{1-b}{2}}\left(n_{\sigma} dx^{\sigma}\right)^{b}\) is invariant under transformations of the \(DISIM_{b}(2)\) group \cite{VSR in Finsler}.
Thus, the general VSR was found to be of Finsler geometry.

There are several reasons why the four Lorentz and CPT--violating models mentioned above could be related with Finsler geometry.
The most fundamental reason is that Finsler geometry gets rid of the quadratic restriction on the spacetime structure \cite{Book by Bao}.
Thus, Finsler geometry is different from Riemann geometry with the quadratic restriction,
although it includes Riemann geometry as a special case.
We refer Finsler geometry as the non--Riemannian Finsler geometry in this paper.
Since Finsler structure gets rid of the quadratic restriction,
the spacetime metric depend on the velocities or momentums of particles propagating in such a spacetime \cite{FSR01,FSR02,FSR03}.
This leads to the MDR for particles and therefore the Lorentz and CPT violations.
On the other hand, Finsler spacetime could depend on certain fixed preferred directions of the spacetime background.
This could be illustrated by the general VSR line element mentioned above,
because this Finsler structure depends on a fixed preferred direction \(n_{\sigma}\) \cite{VSR in Finsler}.
In addition, Finsler spacetime preserves less symmetries than Riemann spacetime.
For instance, the 4D Finsler structure admits
no more than seven Killing vectors \cite{Finsler isometry by Wang,Finsler isometry by Rutz,Finsler isometry LiCM}.
Thus, Finsler spacetime is intrinsically anisotropic.
The Lorentz and CPT violations should appear.
Therefore, Finsler spacetime could be a reasonable platform to describe the Lorentz invariance violation (LIV) and the CPT violation (CPTV).

The LIV and the CPTV may reside in locally Minkowski spacetime \cite{Book by Bao}.
Locally Minkowski spacetime is a generalization of Minkowski spacetime.
Its flag curvature vanishes, and it is a flat Finsler spacetime.
Its metric depends on velocity instead of position.
This property could induce to the LIV and CPTV.
Actually, one could check that the Finsler metrics, related with the four models we mentioned above, belong to locally Minkowski spacetime.
Recently, we have proposed a model of Lorentz--breaking electromagnetic field in such a spacetime \cite{Electromagnetic field in Finsler}.
The LIV was introduced into the Lagrangian of electromagnetic \(U(1)\) field
via the Stueckelberg method \cite{Stueckelberg method01,Stueckelberg method02,Massive photons}.
The Stueckelberg method was employed through replacing the Minkowski metric
with the Finsler metric of locally Minkowski spacetime in our model.
This is similar to the effective metric approach in the SME \cite{Electrodynamics with Lorentz-violating operators of arbitrary dimension}.
We found that the electromagnetic Lagrangian introduced by this method is compatible with the one in the SME at the leading order of the LIV.
In this paper, we generalize our previous method to investigate the non--Abelian gauge fields and matter fields.
We obtain a Lorentz and CPT--breaking extension of the SM.
The LIV and the CPTV properties are discussed miscellaneously in such a model.
In addition, we compare this model with the SME at the leading order.
Both compatibility and otherness are discussed between the SME and our model.

The rest of the paper is arranged as follows.
In section 2, we give a brief review on Finsler spacetime and locally Minkowski spacetime.
Especially, the parity and the time reversal of Finsler structure are presented in locally Minkowski spacetime.
In section 3, we propose Lorentz and CPT--breaking Lagrangians of the gauge fields and matter fields in locally Minkowski spacetime.
In section 4, the LIV and the CPTV are investigated miscellaneously in such an extension of the SM.
In section 5, we compare our proposition with the SME at the leading order of the LIV and the CPTV.
We would derive close relations and distinguishable differences between the SME and our model.
Conclusions and remarks are listed in section 6.

\section{2. Finsler spacetime, parity and time reversal}
\label{Finsler}
Finsler spacetime \cite{Book by Rund,Book by Bao,Book by Shen} is defined on the tangent bundle
\(TM:=\bigcup_{x\in M}T_{x}M\) instead of the manifold \(M\).
Each element of \(TM\) is denoted by \((x,y)\) where \(x\in M\) and \(y:=\frac{d x}{d \tau}\in T_{x}M\).
Finsler geometry originates from the integral of the form \(s=\int^b_a F\left(x, y\right)d\tau\).
The integrand \(F(x,y)\) is called Finsler structure, which is a smooth, positive and positively 1--homogeneous function
defined on the slit tangent bundle \(TM\backslash \{0\}\).
The positive 1--homogeneity means the property \(F(x,\lambda y)=\lambda F(x,y)\) for all \(\lambda>0\).
The Finsler metric is given by
\begin{equation}
\label{Finsler metric}
g_{\mu\nu}(x,y)=\frac{\partial}{\partial y^\mu}\frac{\partial}{\partial y^\nu}\left(\frac{1}{2}F^2\right)\ ,
\end{equation}
which lowers and raises the spacetime indices together with its inverse.
The Finsler metric becomes Riemannian if it does not depend on \(y\).
Thus, Riemann geometry is a special case of Finsler geometry.
In addition, it is obvious that there is no quadratic restriction on the Finsler structure.
The spacetime metric is a function of \(y\).
This may modify the dispersion relations of particles which induces the LIV \cite{DSR in Finsler,FSR01,Electromagnetic field in Finsler}.

Locally Minkowski spacetime \cite{Book by Bao} is a class of Finsler spacetimes with Finsler structures independent of \(x\),
i.e., \(F(x,y)\equiv F(y)\).
Its Finsler metric \(g_{\mu\nu}(y)\) also only depends on \(y\), and its connections and curvatures vanish.
It is the flat Finsler spacetime.
All its tangent spaces are linearly isomorphic to one common linear space.
This means that the physical laws are common at each spacetime position.
A particle moves along a Finsler geodesic \cite{Book by Bao}.
In locally Minkowski spacetime, a free particle moves along a straight line since the connections vanish in the geodesic equations.
The geodesic equations give constant vectors to \(y\),
which means that \(y\) does not dependent on \(x\) \cite{Electromagnetic field in Finsler}.
Locally Minkowski spacetime could be viewed as a straightforward generalization of Minkowski spacetime.

In Minkowski spacetime, the Riemann structure is given by \(F(y)=\left(\eta_{\mu\nu}y^{\mu}y^{\nu}\right)^{\frac{1}{2}}\) where \(\eta_{\mu\nu}=\rm{diag}(+1,-1,-1,-1)\).
It is invariant under the parity (\(P\)) and time reversal (\(T\)) operations.
However, the Finsler metric might change under the \(P\) and \(T\) operations in locally Minkowski spacetime.
One example is the locally Minkowskian Randers structure \(F(y)=\alpha(y)+\beta(y)\)
where \(\alpha=(\eta_{\mu\nu}y^{\mu}y^{\nu})^{\frac{1}{2}}\)
is the Minkowski structure and \(\beta=b_{\mu}y^{\mu}\) denote a 1--form with constant \(b_{\mu}\) \cite{Randers space}.
Obviously, \(\alpha\) is invariant but \(\beta\) changes under the P or T operations.
Correspondingly, the Randers metric could be given as \(g^{\mu\nu}=\eta^{\mu\nu}+\frac{\beta}{\alpha}\left(\eta^{\mu\nu}-\frac{y^{\mu}y^{\nu}}{\alpha^{2}}\right)
+\frac{1}{\alpha}\left(b^{\mu}y^{\nu}+b^{\nu}y^{\mu}\right)+b^{\mu}b^{\nu}\) \cite{Book by Bao}.
One can check that only the first and last terms at the right--hand side are invariant
while the other terms change under the P or T transformations.
The \(P\) and \(T\) violations of Finsler structures imply that the corresponding Finsler spacetimes are asymmetric.
Conversely, the \(P\) and \(T\) violations could originate from departure of the spacetime background from Minkowski spacetime.
Therefore, the CPTV may emerge out in locally Minkowski spacetime.

\section{3. The Lorentz and CPT--breaking Lagrangians}
\label{Lagrangians}
In previous work \cite{Electromagnetic field in Finsler}, we postulated that the LIV resides in locally Minkowski spacetime.
A Lorentz--breaking Lagrangian was proposed for the electromagnetic field
via the Stueckelberg method \cite{Stueckelberg method01,Stueckelberg method02,Massive photons}.
The Minkowski metric was replaced by the Finsler metric of locally Minkowski spacetime in the electromagnetic Lagrangian.
In this way, the principle of relativity is preserved since the Finslerian Lagrangian is covariant under coordinate transformations.
The form of the electromagnetic Lagrangian is similar to the one in the SM, except the Finsler metric.
In addition, the gauge symmetry is still maintained.
In the present paper, we straightforwardly generalize this method to study the Lorentz and CPT--breaking extension of the SM
including the non--Abelian gauge fields and matter fields.

We start our discussions by a brief review on the SM and its conventions.
Especially, the Minkowski metric is extracted in the Lagrangians.
In this way, we could immediately replace it with the Finsler metric in locally Minkowski spacetime later.
In the SM, the interactions between particles are completely determined by the \(SU(3)\times SU(2)\times U(1)\) gauge symmetry.
The full Lagrangian densities for the SM particles could be written as \cite{Book by Peskin}
\begin{eqnarray}
\label{SMlepton}
\mathcal{L}_{lepton}&=&\frac{1}{2}i\overline{L}_{A}\gamma_{\mu}\eta^{\mu\nu}\overleftrightarrow{D}_{\nu}L_{A}
+\frac{1}{2}i\overline{R}_{A}\gamma_{\mu}\eta^{\mu\nu}\overleftrightarrow{D}_{\nu}R_{A}\ ,\\
\label{SMquark}
\mathcal{L}_{quark}&=&\frac{1}{2}i\overline{Q}_{A}\gamma_{\mu}\eta^{\mu\nu}\overleftrightarrow{D}_{\nu}Q_{A}
+\frac{1}{2}i\overline{U}_{A}\gamma_{\mu}\eta^{\mu\nu}\overleftrightarrow{D}_{\nu}U_{A}
+\frac{1}{2}i\overline{D}_{A}\gamma_{\mu}\eta^{\mu\nu}\overleftrightarrow{D}_{\nu}D_{A}\ ,\\
\label{SMYukawa}
\mathcal{L}_{Yukawa}&=&-[(G_{L})_{AB}\overline{L}_{A}\phi R_{B}+(G_{U})_{AB}\overline{Q}_{A}\phi U_{B}+(G_{D})_{AB}\overline{Q}_{A}\phi D_{B}]+h.c.\ ,\\
\label{SMHiggs}
\mathcal{L}_{Higgs}&=&\eta^{\mu\nu}\left(D_{\mu}\phi\right)^{\dagger}D_{\nu}\phi+\mu^{2}\phi^{\dagger}\phi
-\frac{\lambda}{3!}\left(\phi^{\dagger}\phi\right)^{2}\ ,\\
\label{SMgauge}
\mathcal{L}_{gauge}&=&-\frac{1}{2}Tr\left(\eta^{\mu\rho}\eta^{\nu\sigma}G_{\mu\nu}G_{\rho\sigma}\right)
-\frac{1}{2}Tr\left(\eta^{\mu\rho}\eta^{\nu\sigma}W_{\mu\nu}W_{\rho\sigma}\right)
-\frac{1}{4}\eta^{\mu\rho}\eta^{\nu\sigma}B_{\mu\nu}B_{\rho\sigma}\ .
\end{eqnarray}
Throughout of this paper, we employ the conventions as follows.
The lepton and quark multiplets are denoted by
\begin{equation}
\begin{split}
L_{A}=\left(\begin{array}{c}
        \nu_{A} \\
        \ell_{A}
      \end{array}\right)_{L}\ ,\
R_{A}=\left(\ell_{A}\right)_{R}\ ,\
Q_{A}=\left(\begin{array}{c}
        u_{A} \\
        d_{A}
      \end{array}\right)_{L}\ ,\
U_{A}=\left(u_{A}\right)_{R}\ ,\
D_{A}=\left(d_{A}\right)_{R}\ ,\nonumber
\end{split}
\end{equation}
where \(\psi_{L}=\frac{1}{2}(1-\gamma^{5})\psi\) and \(\psi_{R}=\frac{1}{2}(1+\gamma^{5})\psi\)
are, respectively, left-- and right--handed Weyl spinors.
The flavor is labeled by \(A=1,2,3\), which denotes \(\ell_{A}=(e,\mu,\tau)\), \(\nu_{A}=(\nu_{e},\nu_{\mu},\nu_{\tau})\),
\(u_{A}=(u,c,t)\) and \(d_{A}=(d,s,b)\).
The Higgs doublet is \(\varphi\) and its conjugate \(\varphi^{c}\).
We denote \(G_{\mu}\), \(W_{\mu}\) and \(B_{\mu}\) as the SU(3), SU(2) and U(1) gauge fields, respectively.
The related field strengths are given by \(G_{\mu\nu}\), \(W_{\mu\nu}\) and \(B_{\mu\nu}\).
The coupling constants are denoted by \(g_{3}\), \(g\) and \(g'\), respectively.
The Yukawa coupling constants are denoted as \(G_{L}\), \(G_{U}\) and \(G_{D}\).
In addition, \(D_{\mu}\) denotes the covariant derivative and \(A\overleftrightarrow{\partial}_{\mu}B=A\partial_{\mu}B-(\partial_{\mu}A)B\).

The Lorentz--breaking Lagrangian of electromagnetic field has been obtained
via a replacement of the Minkowski metric in the Lorentz--invariant Lagrangian with locally Minkowski metric, i.e.,
\(\eta^{\mu\nu}\longrightarrow g^{\mu\nu}(y)\) \cite{Electromagnetic field in Finsler}.
It was given by
\begin{equation}
\label{Electromagnetic field}
\mathcal{L}_{EM}=-\frac{1}{4}g^{\mu\rho}g^{\nu\sigma}B_{\mu\nu}B_{\rho\sigma}
=-\frac{1}{8}(g^{\mu\rho}g^{\nu\sigma}-g^{\nu\rho}g^{\mu\sigma})B_{\mu\nu}B_{\rho\sigma}\ ,
\end{equation}
where the antisymmetric field strength is \(B_{\mu\nu}=\partial_{\mu}A_{\nu}-\partial_{\nu}A_{\mu}\)
and the 4--potential is \(A_{\mu}(x)\).
In the last step, we had antisymmetrized the indices \(\mu\nu\) and \(\rho\sigma\).
This could be a minimal extension of the quantum electrodynamics (QED), since we did not add any non--standard terms in the Lagrangian.
It is noteworthy that the Lorentz--breaking Lagrangian (\ref{Electromagnetic field}) reduce back into the one of the SM in Minkowski spacetime.

We could straightforwardly obtain the full Lagrangian density of the \(SU(3)\), \(SU(2)\) and \(U(1)\) gauge fields as
\begin{equation}
\label{Finsler Gauge}
\bar{\mathcal{L}}_{gauge}=\frac{1}{2}\left(g^{\mu\rho}g^{\nu\sigma}-g^{\nu\rho}g^{\mu\sigma}\right)
\left[-\frac{1}{2}Tr\left(G_{\mu\nu}G_{\rho\sigma}\right)-\frac{1}{2}Tr\left(W_{\mu\nu}W_{\rho\sigma}\right)
-\frac{1}{4}B_{\mu\nu}B_{\rho\sigma}\right]\ ,
\end{equation}
which result from an analogy to the Lorentz--breaking electromagnetic field.
For the leptons and quarks, their Lagrangians could be given by
\begin{eqnarray}
\label{Finsler Lepton}
\bar{\mathcal{L}}_{lepton}&=&\frac{1}{2}i\overline{L}_{A}\gamma_{\mu}g^{\mu\nu}\overleftrightarrow{D}_{\nu}L_{A}
+\frac{1}{2}i\overline{R}_{A}\gamma_{\mu}g^{\mu\nu}\overleftrightarrow{D}_{\nu}R_{A}\ ,\\
\label{Finsler Quark}
\bar{\mathcal{L}}_{quark}&=&\frac{1}{2}i\overline{Q}_{A}\gamma_{\mu}g^{\mu\nu}\overleftrightarrow{D}_{\nu}Q_{A}
+\frac{1}{2}i\overline{U}_{A}\gamma_{\mu}g^{\mu\nu}\overleftrightarrow{D}_{\nu}U_{A}
+\frac{1}{2}i\overline{D}_{A}\gamma_{\mu}g^{\mu\nu}\overleftrightarrow{D}_{\nu}D_{A}\ ,
\end{eqnarray}
where we directly replace \(\eta\) with locally Minkowski metric \(g\).
The Lagrangian of the Yukawa couplings is as same as the one in the SM, i.e., the equation (\ref{SMYukawa}).
The reason is that the Yukawa couplings do not involve contractions of spacetime indices.
Thus, the full Lagrangian density of the Yukawa couplings for the matter fields is given by
\begin{equation}
\label{Finsler Yukawa}
\bar{\mathcal{L}}_{Yukawa}=\mathcal{L}_{Yukawa}\ .
\end{equation}
For the Higgs sector, we could obtain its Lagrangian as
\begin{equation}
\label{Finsler Higgs}
\bar{\mathcal{L}}_{Higgs}=g^{\mu\nu}\left(D_{\mu}\phi\right)^{\dagger}D_{\nu}\phi+\mu^{2}\phi^{\dagger}\phi
-\frac{\lambda}{3!}\left(\phi^{\dagger}\phi\right)^{2}\ ,
\end{equation}
where we once again directly replace the Minkowski metric with the locally Minkowski metric.
In this paper, the bars over \(\mathcal{L}\) denote the Lagrangian densities of fields in locally Minkowski spacetime.

The five equations (\ref{Finsler Gauge})--(\ref{Finsler Higgs}) constitute a complete set of Lorentz--breaking Lagrangians
of the gauge and matter fields in locally Minkowski spacetime.
The Lagrangians are obtained by replacing the Minkowski metric in the SM ones with the locally Minkowski metric.
This implies that these Lorentz--breaking Lagrangians could reduce back to
those Lorentz--invariant Lagrangians in the SM at the Minkowski limit.
In addition, these Lagrangians are invariant under coordinate transformations since they are spacetime scalars.
On the other hand, the CPTV may emerge out in locally Minkowski spacetime.
The reason is that the locally Minkowski metric \(g_{\mu\nu}\) could include both the CPT--even and CPT--odd parts,
such as the Randers metric mentioned in last section.
The coupling constants induced by the LIV and CPTV are constant in the SME.
The SME terms involving \((c_{L,Q})^{\mu\nu}\), \((k_{\phi\phi})^{\mu\nu}\) and \((k_{G,W,B})^{\mu\nu\rho\sigma}\) are CPT--even \cite{SME02}.
However, the corresponding terms get rid of this property in our model.
They could consist of both the CPT--even and CPT--odd parts simultaneously, which are determined completely by the locally Minkowski metric.
We will discuss this issue in the next sections.

\section{4. The LIV and the CPTV}
\label{LIV and CPTV}
Physical observations have provided severe constraints on possible LIV and CPTV \cite{Data tables for Lorentz and CPT violation}.
These constraints reveal that the LIV and CPTV should be tiny enough to escape the present observable sensitivity.
Thus, we could study the perturbative implications of the equation set (\ref{Finsler Gauge})---(\ref{Finsler Higgs}).
In this section, we expand the locally Minkowski metric around the Minkowski metric to the leading order of the LIV.
In other word, the locally Minkowski metric can now be written as
\begin{equation}
\label{Finsler metric LO}
g^{\mu\nu}=\eta^{\mu\nu}+h^{\mu\nu}+\mathcal{O}(h^{2})\ .
\end{equation}
All the possible LIV and CPTV result completely from the perturbative term \(h^{\mu\nu}\) in locally Minkowski spacetime.
Correspondingly, we could obtain the perturbative expansions of
the Lorentz--breaking Lagrangians (\ref{Finsler Gauge})---(\ref{Finsler Higgs}).
To the leading order, the Lorentz--breaking perturbations \(\delta\mathcal{L}=\bar{\mathcal{L}}-\mathcal{L}\) of the Lagrangians are given by
\begin{eqnarray}
\label{GaugeLO}
\delta\mathcal{L}_{gauge}&=&\frac{1}{2}\left(\eta^{\mu\rho}h^{\nu\sigma}-\eta^{\nu\rho}h^{\mu\sigma}
-\eta^{\mu\sigma}h^{\nu\rho}+\eta^{\nu\sigma}h^{\mu\rho}\right)\left[-\frac{1}{2}Tr(G_{\mu\nu}G_{\rho\sigma})-\frac{1}{2}Tr(W_{\mu\nu}W_{\rho\sigma})
-\frac{1}{4}B_{\mu\nu}B_{\rho\sigma}\right]\ ,\\
\label{LeptonLO}
\delta\mathcal{L}_{lepton}&=&
\frac{1}{2}ih^{\mu\nu}\overline{L}_{A}\gamma_{\mu}\overleftrightarrow{D}_{\nu}L_{A}
+\frac{1}{2}ih^{\mu\nu}\overline{R}_{A}\gamma_{\mu}\overleftrightarrow{D}_{\nu}R_{A}\ ,\\
\label{QuarkLO}
\delta\mathcal{L}_{quark}&=&
\frac{1}{2}ih^{\mu\nu}\overline{Q}_{A}\gamma_{\mu}\overleftrightarrow{D}_{\nu}Q_{A}
+\frac{1}{2}ih^{\mu\nu}\overline{U}_{A}\gamma_{\mu}\overleftrightarrow{D}_{\nu}U_{A}
+\frac{1}{2}ih^{\mu\nu}\overline{D}_{A}\gamma_{\mu}\overleftrightarrow{D}_{\nu}D_{A}
\ ,\\
\label{YukawaLO}
\delta\mathcal{L}_{Yukawa}&=&0\ ,\\
\label{HiggsLO}
\delta\mathcal{L}_{Higgs}&=&h^{\mu\nu}\left(D_{\mu}\phi\right)^{\dagger}D_{\nu}\phi\ .
\end{eqnarray}
The coupling constants in the above equations carry spacetime indices, which reflects the LIV properties.
In addition, the LIV and the CPTV do not affect the Yukawa coupling terms which generate mass of particles.
In the quantum field theory (QFT), causality requires every particle having a relating antiparticle
with opposite quantum numbers and same mass \cite{Book by Peskin}.
Therefore, causality is preserved in our model even through the Lorentz invariance and the CPT symmetry are broken.

We notice that all the LIV effects in Eq.(\ref{GaugeLO})--(\ref{HiggsLO}) result from the metric perturbation \(h^{\mu\nu}\)
and their CPT properties are determined completely by the CPT properties of \(h^{\mu\nu}\).
The LIV--induced coupling constants are similar for the matter fields and the Higgs field.
All of them are characterized directly by the metric perturbation \(h^{\mu\nu}\).
For the gauge sectors, their LIV--induced coupling constants are one common multiplication of
\(\eta^{\mu\rho}h^{\nu\sigma}\) together with its anti-symmetrizations of spacetime indices,
i.e.,
\begin{equation}
\label{k}
k^{\mu\nu\rho\sigma}\equiv\frac{1}{2}\left(\eta^{\mu\rho}h^{\nu\sigma}-\eta^{\nu\rho}h^{\mu\sigma}
-\eta^{\mu\sigma}h^{\nu\rho}+\eta^{\nu\sigma}h^{\mu\rho}\right)\ .
\end{equation}
They have the symmetries of the Riemann curvature tensor which have nineteen components.
However, there are only ten independent components for \(h^{\mu\nu}\).
Thus, \(k^{\mu\nu\rho\sigma}\) have not more than ten components independent.
In this sense, the metric perturbation \(h^{\mu\nu}\) totally determines all the possible LIV.
This result is consistent with the kinematical predictions \cite{DSR in Finsler,FSR03,Randers SR Chang01,Randers SR Chang02}.
In section 2, we have shown that the locally Minkowski line element, such as the Randers one, may not be invariant
under the parity and the time reversal operations.
The metric perturbation \(h^{\mu\nu}\) could consist of the CPT--even and the CPT--odd parts.
It completely determines the CPT--properties of the Lorentz--breaking terms in the Lagrangians.
For instance, we consider the locally Minkowskian Randers structure once again.
The Randers metric perturbation could be given as \(h^{\mu\nu}=\frac{\beta}{\alpha}\left(\eta^{\mu\nu}-\frac{y^{\mu}y^{\nu}}{\alpha^{2}}\right)
+\frac{1}{\alpha}\left(b^{\mu}y^{\nu}+b^{\nu}y^{\mu}\right)+b^{\mu}b^{\nu}\) \cite{Book by Bao}.
Based on discussions in section 2, one could check that the first two terms are CPT--odd
and the third one is CPT--even on the right hand side.
Therefore, \(h^{\mu\nu}\) could have hybrid CPT property.

\section{5. Comparison with the SME}
\label{Comparison}
It is helpful to make a comparison between the SME and this Finslerian SM.
The SME has been investigated extensively and the LIV and the CPTV are well understood phenomenologically.
By comparing the Finslerian SM with the SME, we could find relations and differences between them.
This benefits understanding the Finslerian SM from phenomenologies and experimental implications.
In locally Minkowski spacetime, the Lorentz--breaking Lagrangian of the electromagnetic field has been shown to be
compatible with the one in the SME at the leading order \cite{Electromagnetic field in Finsler}.
In this section, we would show that the Lorentz and CPT--breaking Lagrangians (\ref{GaugeLO})--(\ref{HiggsLO})
are also formally compatible with those in the SME for the matter fields and the non--Abelian fields.
Certainly, we also notice distinguishable differences between these two models.

In the SME, the LIV and CPTV perturbations of Lagrangian for the gauge fields are given by \cite{SME02}
\begin{eqnarray}
\label{GaugeSME CPT-even}
\delta\mathcal{L}_{gauge}^{CPT-even}&=&-\frac{1}{2}(k_{G})^{\mu\nu\rho\sigma}Tr(G_{\mu\nu}G_{\rho\sigma})
-\frac{1}{2}(k_{W})^{\mu\nu\rho\sigma}Tr(W_{\mu\nu}W_{\rho\sigma})
-\frac{1}{4}(k_{B})^{\mu\nu\rho\sigma}B_{\mu\nu}B_{\rho\sigma}\ ,\\
\label{GaugeSME CPT-odd}
\delta\mathcal{L}_{gauge}^{CPT-odd}&=&(k_{3})_{\mu}\epsilon^{\mu\nu\rho\sigma}Tr(G_{\nu}G_{\rho\sigma}+\frac{2}{3}i g_{3} G_{\nu}G_{\rho}G_{\sigma})
+(k_{2})_{\mu}\epsilon^{\mu\nu\rho\sigma}Tr(W_{\nu}W_{\rho\sigma}+\frac{2}{3}i g W_{\nu}W_{\rho}W_{\sigma})\nonumber\\
& &+(k_{1})_{\mu}\epsilon^{\mu\nu\rho\sigma}B_{\nu}B_{\rho\sigma}+(k_{0})^{\mu}B_{\mu}\ .
\end{eqnarray}
By comparing (\ref{GaugeSME CPT-even}) (\ref{GaugeSME CPT-odd}) with (\ref{GaugeLO}),
we find two formal relations \((k_{G,W,B})^{\mu\nu\rho\sigma}=k^{\mu\nu\rho\sigma}\) and \((k_{i})_{\mu}=0\) for \(i=0,1,2,3\).
The latter relation reveals that the SME--like CPT--odd terms would not appear in the Finslerian SM.
Only the SME--like CPT--even terms could appear according to the former relation.
However, \(k\) comprises \(\eta\) and \(h(y)\) according to (\ref{k}).
Thus, the terms involving \(k^{\mu\nu\rho\sigma}\) could be CPT--even, or CPT--odd or even CPT--hybrid in the Finslerian SM.
This prediction is different from that in the SME.
In addition, the LIV and CPTV properties are similar for three gauge fields.
Furthermore, \(k\) is a function of the velocity \(y\) of a particle.
However, there could exist different velocities for particles.
Thus the Lorentz--breaking effects could be observable.
We will see in the following that these predictions still hold for the matter fields.

For the leptons and quarks, the perturbations of Lagrangians are given as \cite{SME02}
\begin{eqnarray}
\label{LeptonSME_CPT-even}
\delta\mathcal{L}_{lepton}^{CPT-even}&=&\frac{1}{2}i(c_{L})^{\mu\nu}_{AB}\overline{L}_{A}\gamma_{\mu}\overleftrightarrow{D}_{\nu}L_{B}
+\frac{1}{2}i(c_{R})^{\mu\nu}_{AB}\overline{R}_{A}\gamma_{\mu}\overleftrightarrow{D}_{\nu}R_{B}\ ,\\
\label{LeptonSME CPT-odd}
\delta\mathcal{L}_{lepton}^{CPT-odd}&=&-(a_{L})^{\mu}_{AB}\overline{L}_{A}\gamma_{\mu}L_{B}
-(a_{R})^{\mu}_{AB}\overline{R}_{A}\gamma_{\mu}R_{B}\ ,\\
\label{QuarkSME CPT-even}
\delta\mathcal{L}_{quark}^{CPT-even}&=&\frac{1}{2}i(c_{Q})^{\mu\nu}_{AB}\overline{Q}_{A}\gamma_{\mu}\overleftrightarrow{D}_{\nu}Q_{B}
+\frac{1}{2}i(c_{U})^{\mu\nu}_{AB}\overline{U}_{A}\gamma_{\mu}\overleftrightarrow{D}_{\nu}U_{B}
+\frac{1}{2}i(c_{D})^{\mu\nu}_{AB}\overline{D}_{A}\gamma_{\mu}\overleftrightarrow{D}_{\nu}D_{B}\ ,\\
\label{QuarkSME CPT-odd}
\delta\mathcal{L}_{quark}^{CPT-odd}&=&-(a_{Q})^{\mu}_{AB}\overline{Q}_{A}\gamma_{\mu}Q_{B}
-(a_{U})^{\mu}_{AB}\overline{U}_{A}\gamma_{\mu}U_{B}
-(a_{D})^{\mu}_{AB}\overline{D}_{A}\gamma_{\mu}D_{B}\ .
\end{eqnarray}
Again we find \((a_{L,R,Q,U,D})^{\mu}_{AB}=0\) for the SME CPT--odd terms
and \((c_{L,R,Q,U,D})^{\mu\nu}_{AB}=h^{\mu\nu}\delta_{AB}\) for the SME CPT--even terms by comparing to (\ref{LeptonLO}) (\ref{QuarkLO}).
Only the symmetric part could exist for \(c\).
In addition, the LIV and the CPTV would not mix different flavors since there exists \(\delta_{AB}\).

There exists only CPT--even Lorentz--breaking terms for the Yukawa couplings in the SME \cite{SME02}
\begin{eqnarray}
\label{YukawaSME CPT-even}
\delta\mathcal{L}_{Yukawa}^{CPT-even}&=&-\frac{1}{2}\left[(H_{L})_{\mu\nu AB}\overline{L}_{A}\phi\sigma^{\mu\nu}R_{B}
+(H_{U})_{\mu\nu AB}\overline{Q}_{A}\phi\sigma^{\mu\nu}U_{B}
+(H_{D})_{\mu\nu AB}\overline{Q}_{A}\phi\sigma^{\mu\nu}D_{B}\right]+h.c.\ .
\end{eqnarray}
However, the above coupling constants vanish, i.e., \(H_{L,U,D}=0\),
since the LIV and the CPTV do not affect the Yukawa couplings in (\ref{YukawaLO}).
On the other hand, the Higgs field acquires the same vev as that in the SM, which will be discussed below.
Thus, the mass of matter particles would not be influenced by the LIV and the CPTV.

For the Higgs sector, the LIV and CPTV terms of its Lagrangian are \cite{SME02}
\begin{eqnarray}
\label{HiggsSME CPT-even}
\delta\mathcal{L}_{Higgs}^{CPT-even}&=&\frac{1}{2}(k_{\phi\phi})^{\mu\nu}(D_{\mu}\phi)^{\dagger}D_{\nu}\phi+h.c.
-\frac{1}{2}(k_{\phi B})^{\mu\nu}\phi^{\dagger}B_{\mu\nu}\phi
-\frac{1}{2}(k_{\phi W})^{\mu\nu}\phi^{\dagger}W_{\mu\nu}\phi\ ,\\
\label{HiggsSME CPT-odd}
\delta\mathcal{L}_{Higgs}^{CPT-odd}&=&i(k_{\phi})^{\mu}\phi^{\dagger}D_{\mu}\phi+h.c.\ .
\end{eqnarray}
By comparing to (\ref{HiggsLO}), we obtain relations \((k_{\phi\phi}+k_{\phi\phi}^{\ast})^{\mu\nu}=2h^{\mu\nu}\)
and \(k_{\phi}=k_{\phi B}=k_{\phi W}=0\).
In the SME, the spontaneous breaking of electroweak symmetry gives the nonzero vev
\(\langle Z_{\mu}^{0}\rangle=\frac{1}{q}\rm{sin}2\theta_{W}(\rm{Re}(\eta+k_{\phi\phi}))^{-1}_{\mu\nu}k^{\nu}_{\phi}\) to \(Z_{\mu}^{0}\)
and \(\langle \varphi\rangle=a(1-\mu^{-2}(\rm{Re}(\eta+k_{\phi\phi}))^{-1}_{\mu\nu}k_{\phi}^{\mu}k_{\phi}^{\nu})^{1/2}\)
to the Higgs field \cite{SME02}. Here, \(q=g\rm{sin}\theta_{W}\) denotes U(1) charge and \(a=(6\mu^{2}/\lambda)^{1/2}\).
However, the vev of \(Z_{\mu}^{0}\) would vanish in the Finslerian SM for \(k_{\phi}=0\)
and the Higgs field would acquire the same vev as that in the SM.

From the above discussions, we find that the Finslerian SM could be formally related to the SME framework at the leading order.
The LIV--induced coupling constants could be combined to represent the SME coupling constants.
Since there are only ten independent components of \(h^{\mu\nu}\),
the parameter space of the Finslerian SM is severely squeezed.
Certainly, there are also many differences between the SME and our model.
The most significant difference is that the LIV--induced coupling constants are constant in the SME
while they could depend on momentums of particles in the Finslerian SM.
In the SME, the so--called CPT--even terms are really CPT--even.
However, the corresponding terms could be CPT--even, or CPT--odd, or even CPT--hybrid.
This is determined completely by the locally Minkowski metric.
This prediction might help to distinguish the Finslerian SM from the SME in forthcoming experiments.
The other significant difference is that the Higgs field has a modified vev in the SME
while it remains the same as that of the SM in the Finslerian SM.
Thus, the mass of other particles would remain unchanged.

To demonstrate the above discussions clearly, we consider the effective Hamiltonian \(h_{eff}\)
which describes flavor neutrino propagation and oscillations.
In the SME, this effective Hamiltonian could be given by \cite{LIV and CPTV in neutrinos}
\begin{equation}
\label{effective Hamiltonian}
\left(h_{eff}\right)_{AB}=|\vec{p}|\delta_{AB}+\frac{1}{2|\vec{p}|}\left(m^{2}\right)_{AB}
+\frac{1}{|\vec{p}|}\left[(a_{L})^{\mu}p_{\mu}-(c_{L})^{\mu\nu}p_{\mu}p_{\nu}\right]_{AB}\ .
\end{equation}
It is invariant under CPT unless \(a_{L}\) vanishes.
For two--flavor system \(\nu_{\mu}\rightarrow\nu_{\tau}\) with maximal mixing,
the correction to \(\nu_{\mu}\) disappearance probability is \cite{Long-baseline neutrino experiments as tests for Lorentz violation}
\begin{equation}
P^{(1)}_{\nu_{\mu}\rightarrow\nu_{\tau}}\simeq \rm{Re} \left(\frac{1}{|\vec{p}|}\left[(a_{L})^{\mu}p_{\mu}-(c_{L})^{\mu\nu}p_{\mu}p_{\nu}\right]_{\nu_{\mu}\nu_{\tau}}\right)
~L~\rm{sin}\left(\frac{\Delta m^{2}_{\nu_{\mu}\nu_{\tau}}L}{2E}\right)\ ,
\end{equation}
where \(L\) denotes a baseline.
The \(a_{L}\) term controls the LIV and CPTV while the \(c_{L}\) term controls the LIV only.
When \(a_{L}\) vanishes, the CPT is preserved, i.e., \(P^{(1)}_{\nu_{\mu}\rightarrow\nu_{\tau}}=P^{(1)}_{\bar{\nu}_{\tau}\rightarrow\bar{\nu}_{\mu}}\).
In locally Minkowski spacetime, the situation \(a_{L}=0\) is always preserved.
We have obtained a formal relation \(\left(c_{L}\right)^{\mu\nu}=h^{\mu\nu}\).
However, \(h^{\mu\nu}\) is not constant now.
It depends on velocity of neutrinos.
Its CPT property determines the CPT property of the effective Hamiltonian.
For example, the Randers metric occupies both the CPT--even and CPT--odd parts as mentioned in last section.
The effective Hamiltonian could be given as
\begin{equation}
\left(\bar{h}_{eff}\right)_{AB}=|\vec{p}|\delta_{AB}+\frac{1}{2|\vec{p}|}\left(m^{2}\right)_{AB}
+\frac{1}{|\vec{p}|}\left[-h^{\mu\nu}_{even}p_{\mu}p_{\nu}-h^{\mu\nu}_{odd}p_{\mu}p_{\nu}\right]\ .
\end{equation}
We have decomposed \(h^{\mu\nu}=h^{\mu\nu}_{even}+h^{\mu\nu}_{odd}\),
where \(h^{\mu\nu}_{even}\) and \(h^{\mu\nu}_{odd}\) denote the CPT--even and CPT--odd parts, respectively.
Thus the Finslerian terms corresponding to the SME \(c_{L}\) term could occupy hybrid CPT behavior.
For the above \(\nu_{\mu}\rightarrow\nu_{\tau}\) system, the correction to \(\nu_{\mu}\) disappearance probability is
\begin{equation}
\bar{P}^{(1)}_{\nu_{\mu}\rightarrow\nu_{\tau}}\simeq \left(\frac{1}{|\vec{p}|}\left[-h^{\mu\nu}_{even}p_{\mu}p_{\nu}-h^{\mu\nu}_{odd}p_{\mu}p_{\nu}\right]_{\nu_{\mu}\nu_{\tau}}\right)
~L~\rm{sin}\left(\frac{\Delta m^{2}_{\nu_{\mu}\nu_{\tau}}L}{2E}\right)\ .
\end{equation}
We have shown that both the terms correspond to the \(c_{L}\) term of the SME.
However, the \(h^{\mu\nu}_{odd}\) term controls the LIV and CPTV while the \(h^{\mu\nu}_{even}\) term controls only the LIV in this case.
We find the CPTV, i.e., \(\bar{P}^{(1)}_{\nu_{\mu}\rightarrow\nu_{\tau}}\neq \bar{P}^{(1)}_{\bar{\nu}_{\tau}\rightarrow\bar{\nu}_{\mu}}\).
This prediction is different from that in the SME.

\section{6. Conclusions and remarks}
\label{Conclusions}
In this paper, we proposed a Finsler geometric framework for investigating
extension of the SM of particle physics with the LIV and the CPTV effects.
The Lorentz and CPT--breaking Lagrangians of the SM fields were presented explicitly in locally Minkowski spacetime.
The LIV and the CPTV were introduced into the Lagrangians of fields
by replacing the Minkowski metric with the locally Minkowski metric.
The obtained Lagrangians could reduce back to those in the SM at the Minkowski limit.
The principle of relativity and the gauge symmetries are still preserved for this model.
The LIV together with the CPTV was found to have a possible geometric origin of spacetime structure deviating from the Minkowski one.
In addition, the CPTV is completely determined by behaviors of locally Minkowski structure under the parity and time reversal.

At the leading order, the Finslerian model could be formally related with the SME.
The LIV--induced coupling constants are characterized by the metric deviation from the Minkowski metric in this model.
They could be combined into the corresponding SME coupling constants.
However, the parameter space is squeezed severely for these SME coupling constants.
The reason is that only ten independent components exist in the metric deviation.
We discussed the most significant difference between the SME and the Finslerian SM.
In locally Minkowski spacetime, the spacetime metric could be a function of momentums of particles.
It, as well as its deviation, might be CPT--even, or CPT--odd or even CPT--hybrid.
Therefore, the relating terms of the Lagrangians would hold the same CPT properties.
This prediction is different from that of the SME.
It may help to distinguish this locally Minkowskian model from the SME in forthcoming experiments.

\vspace{0.4 cm}

\begin{acknowledgments}
We thank useful discussions with Y. Jiang, M.-H. Li, X. Li, H.-N. Lin, J.-P. Dai, D.-N. Li, and X.-G. Wu.
This work is supported by the National Natural Science Fund of China under Grant No. 11075166.
\end{acknowledgments}


\begin{thebibliography}{999}

\bibitem{SME01}D. Colladay and V. A. Kostelecky, Phys. Rev. D {\bf 55}, 6760 (1997).
\bibitem{SME02}D. Colladay and V. A. Kostelecky, Phys. Rev. D {\bf 58}, 116002 (1998).

\bibitem{SBLV}V. A. Kostelecky and S. Samuel, Phys. Rev. D {\bf 39}, 683 (1989).

\bibitem{recoiling D-branes01}J. R. Ellis, N. E. Mavromatos and D. V. Nanopoulos, Gen. Rel. Grav. {\bf 32}, 127 (2000).
\bibitem{recoiling D-branes02}J. R. Ellis, N. E. Mavromatos and D. V. Nanopoulos, Phys. Rev. D {\bf 61}, 027503 (1999).
\bibitem{recoiling D-branes03}J. R. Ellis, N. E. Mavromatos and D. V. Nanopoulos, Phys. Rev. D {\bf 62}, 084019 (2000).
\bibitem{spacetime foam01}J. Alfaro, H. A. Morales-Tecotl and L. F. Urrutia, Phys. Rev. Lett. {\bf 84}, 2318 (2000).

\bibitem{DSR01}G. Amelino-Camelia, Phys. Lett. B {\bf 510}, 255 (2001).
\bibitem{DSR02}G. Amelino-Camelia, Int. J. Mod. Phys. D {\bf 11}, 35 (2002).
\bibitem{DSR03}G. Amelino-Camelia, Nature {\bf 418}, 34 (2002).
\bibitem{DSR04}J. Magueijo and L. Smolin, Phys. Rev. Lett. {\bf 88}, 190403 (2002).
\bibitem{DSR05}J. Magueijo and L. Smolin, Phys. Rev. D {\bf 67}, 044017 (2003).

\bibitem{MDR in DSR}S. Ghosh and P. Pal, Phys. Rev. D {\bf 75}, 105021 (2007).


\bibitem{VSR}A. G. Cohen and S. L. Glashow, Phys. Rev. Lett. {\bf 97}, 021601 (2006).

\bibitem{Book by Rund}H. Rund, {\it The Differential Geometry of Finsler Spaces}, Springer, Berlin, 1959.
\bibitem{Book by Bao}D. Bao, S. S. Chern, and Z. Shen, {\it An Introduction to Riemann--Finsler Geometry},
        Graduate Texts in Mathmatics {\bf 200}, Springer, New York, 2000.
\bibitem{Book by Shen}Z. Shen, {\it Lectures on Finsler Geometry}, World Scientific, Singapore, 2001.

\bibitem{SME Bogoslovsky01}G.Yu. Bogoslovsky, Phys. Lett. A {\bf 350}, 5 (2006).
\bibitem{SME Bogoslovsky02}G.Yu. Bogoslovsky, SIGMA {\bf 1}, 017 (2005).
\bibitem{Kostelecky_Finsler}V. A. Kostelecky, Phys. Lett. B {\bf 701}, 137 (2011).

\bibitem{DSR in Finsler}F. Girelli, S. Liberati and L. Sindoni, Phys. Rev. D {\bf 75}, 064015 (2007).

\bibitem{Rainbow metric}J. Magueijo and L. Smolin, Class. Quant. Grav.{\bf 21}, 1725 (2004).

\bibitem{VSR in Finsler}G.W. Gibbons, J. Gomis and C. N. Pope, Phys. Rev. D {\bf 76}, 081701 (2007).

\bibitem{FSR01}Z. Chang, X. Li and S. Wang, Mod. Phys. Lett. A {\bf 27}, 1250058 (2012).
\bibitem{FSR02}Z. Chang, X. Li and S. Wang, Phys. Lett. B {\bf 710}, 430 (2012).
\bibitem{FSR03}Z. Chang, X. Li and S. Wang, arXiv:1201.1368.

\bibitem{Finsler isometry LiCM}X. Li and Z. Chang, arXiv:1010.2020.
\bibitem{Finsler isometry by Wang}H. C. Wang, J. London Math. Soc. {\bf s1-22} (1), 5-9 (1947).
\bibitem{Finsler isometry by Rutz}S. F. Rutz, Contemp. Math. {\bf 169}, 289 (1996).
%\bibitem{Finsler isometry}S. Deng and Z. Hou, Pac. J. Math. {\bf 207}, 1 (2002).

\bibitem{Electromagnetic field in Finsler}Z. Chang and S. Wang, arXiv:1204.2478, accepted by Eur. Phys. J. C.

\bibitem{Stueckelberg method01}E. C. G. Stueckelberg, Helv. Phys. Acta {\bf 11}, 225 (1938).
\bibitem{Stueckelberg method02}E. C. G. Stueckelberg, Helv. Phys. Acta {\bf 11}, 299 (1938).
\bibitem{Massive photons}M. Cambiaso, R. Lehnert and R. Potting, Phys. Rev. D {\bf 85}, 085023 (2012).

\bibitem{Electrodynamics with Lorentz-violating operators of arbitrary dimension}V. A. Kostelecky and M. Mewes, Phys.Rev.D {\bf 80}, 015020 (2009).

\bibitem{Randers space}G. Randers, Phys, Rev. {\bf 59}, 195(1941).

\bibitem{Book by Peskin}M. E. Peskin and D. V. Schroeder, {\it An introduction to Quantum Field Theory}, Westview, 1995.

\bibitem{Data tables for Lorentz and CPT violation}V. A. Kostelecky and N. Russell, Rev. Mod. Phys. {\bf 83}, 11 (2011).

\bibitem{Randers SR Chang01}Z. Chang and X. Li, Chinese Phys. C {\bf 33}, 626 (2009).
\bibitem{Randers SR Chang02}Z. Chang and X. Li, Phy. Lett. B {\bf 663}, 103 (2008).

\bibitem{LIV and CPTV in neutrinos}V.A. Kostelecky and M. Mewes, Phys. Rev. D {\bf 69}, 016005 (2004).

\bibitem{Long-baseline neutrino experiments as tests for Lorentz violation}J. S. Diaz, arXiv:0909.5360 [hep-ph], IUHET 532 (2009).

\end{thebibliography}
\end{document}